\documentclass[10pt,conference]{IEEEtran}
\usepackage{maxim}
\usepackage{verbatim}
\usepackage{graphicx}

\def\cA{{\cal A}}
\def\cF{{\cal F}}
\def\cL{{\cal L}}
\def\cM{{\cal M}}
\def\cN{{\cal N}}
\def\cP{{\cal P}}
\def\cX{{\cal X}}
\def\cY{{\cal Y}}
\def\cZ{{\cal Z}}

\def\bbD{{\mathbb D}}

\def\wh#1{\widehat{#1}}

\renewcommand{\theequation}{\arabic{equation}}
\begin{document}
\bibliographystyle{IEEEtran}

\title{Learning From Compressed Observations}

\author{
\authorblockN{Maxim Raginsky}
\authorblockA{Beckman Institute and University of Illinois \\
405 N Mathews Ave, Urbana, IL 61801 \\
maxim@uiuc.edu}
}

\maketitle

\begin{abstract} The problem of statistical learning is to construct a predictor of a random variable $Y$ as a function of a related random variable $X$ on the basis of an i.i.d.~training sample from the joint distribution of $(X,Y)$. Allowable predictors are drawn from some specified class, and the goal is to approach asymptotically the performance (expected loss) of the best predictor in the class. We consider the setting in which one has perfect observation of the $X$-part of the sample, while the $Y$-part has to be communicated at some finite bit rate. The encoding of the $Y$-values is allowed to depend on the $X$-values. Under suitable regularity conditions on the admissible predictors, the underlying family of probability distributions and the loss function, we give an information-theoretic characterization of achievable predictor performance in terms of conditional distortion-rate functions. The ideas are illustrated on the example of nonparametric regression in Gaussian noise.\end{abstract}

\section{Introduction and problem statement}
\label{sec:intro}

Let $X$ and $Y$ be jointly distributed random variables, where $X$ takes values in an {\em input space} $\cX$ and $Y$ takes values in an {\em output space} $\cY$. The problem of statistical learning is about constructing an accurate predictor of $Y$ as a function of $X$ on the basis of some number of independent copies of $(X,Y)$, often with very little or no prior knowledge of the underlying distribution. A very general decision-theoretic framework for learning was proposed by Haussler \cite{Hau92}. In a slightly simplified form it goes as follows. Let $\cP$ be a family of probability distributions on $\cZ = \cX \times \cY$. Each member $P$ of $\cP$ represents a possible  relationship between $X$ and $Y$. Also given are a {\em loss function} $\map{\ell}{\cY \times \cY}{\R^+}$ and a set $\cF$ of functions ({\em hypotheses}) from $\cX$ into $\cY$. For any $f \in \cF$ and any $P \in \cP$ we have the {\em expected loss} (or {\em risk})
$$
L(f,P) = \E\ell(f(X),Y) \equiv \int_\cZ \ell(f(x),y)dP(x,y),
%\label{eq:expected_loss}
$$
which expresses quantitatively the average performance of $f$ as a predictor of $Y$ from $X$ when $(X,Y) \sim P$. Let us define the minimum expected loss
$$
L^*(\cF,P) \deq \inf_{f \in \cF} L(f,P)
%\label{eq:min_expected_loss}
$$
and assume that the infimum is achieved by some $f^* \in \cF$. Then $f^*$ is the best predictor of $Y$ from $X$ in the hypothesis class $\cF$ when $(X,Y) \sim P$. The problem of statistical learning is to construct, for each $n$, an approximation to $f^*$ on the basis of a {\em training sequence} $\{Z_i\}^n_{i=1}$, where $Z_i = (X_i,Y_i)$ are i.i.d.~according to $P$, such that this approximation gets better and better as the sample size $n$ tends to infinity. This formulation of the learning problem is referred to as {\em agnostic} (or {\em model-free}) learning, reflecting the fact that typically only minimal assumptions are made on the causal relation between $X$ and $Y$ and on the capability of the hypotheses in $\cF$ to capture this relation. It is general enough to cover such problems as classification, regression and density estimation.

Formally, a {\em learning algorithm} (or {\em learner}, for short) is a sequence $\{ \wh{f}_n \}^\infty_{n=1}$ of maps $\map{\wh{f}_n}{\cZ^n \times \cX}{\cY}$, such that $\wh{f}_n(Z^n,\cdot) \in \cF$ for all $n$ and all $Z^n \in \cZ^n$. Let $Z = (X,Y) \sim P$ be independent of the training sequence $Z^n$. The main quantity of interest is the {\em generalization error} of the learner,
\begin{eqnarray*}
L(\wh{f}_n, P) &\deq& \E \Big[ \ell \big( \wh{f}_n(Z^n,X),Y \big) \Big | Z^n \Big] \\
&\equiv& \int_\cZ \ell(\wh{f}_n(Z^n,x),y)dP(x,y).
%\label{eq:generalization_error}
\end{eqnarray*}
The generalization error is a random variable, as it depends on the training sequence $Z^n$. One is chiefly interested in the asymptotic probabilistic behavior of the {\em excess loss} $L(\wh{f}_n, P) - L^*(\cF,P)$ as $n \to \infty$. (Clearly, $L(\wh{f}_n, P) \ge L^*(\cF,P)$ for every $n$.) Under suitable conditions on the loss function $\ell$, the hypothesis class $\cF$, and the underlying family $\cP$ of probability distributions, one can show that there exist learning algorithms which not only {\em generalize}, i.e., $\E L(\wh{f}_n, P) \to L^*(\cF,P)$ as $n \to \infty$ for every $P \in \cP$ (which is the least one could ask for), but are also {\em probably approximately correct} (PAC), i.e.
\begin{equation}
\lim_{n \to \infty} P \left( Z^n : L(\wh{f}_n, P) > L^*(\cF,P) + \epsilon \right) = 0
\label{eq:universal_consistency}
\end{equation}
for every $\epsilon > 0$ and every $P \in \cP$. (See, e.g., Vidyasagar \cite{Vid03}.)

\begin{figure}
\includegraphics[width=\columnwidth]{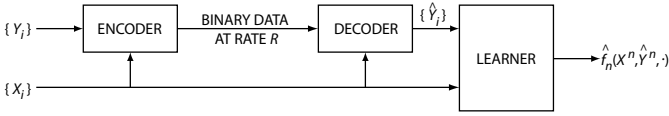}
\caption{The set-up for learning from compressed data with side information.}
\label{fig:rate_limited_learning}
\end{figure}

This formulation assumes that the training data are available to the learner with arbitrary precision. This assumption may not always hold, however. For example, the location at which the training data are gathered may be geographically separated from the location where the learning actually takes place. Therefore, the training data may have to be communicated to the learner over a channel of finite capacity. In that case, the learner will see only a quantized version of the training data, and must be able to cope with this to the extent allowed by the fundamental limitations imposed by rate-distortion theory. In this paper, we consider a special case of such learning under rate constraints, when the learner has perfect observation of the input part $X^n = (X_1,\ldots,X_n)$ of the training sequence, while the output part $Y^n = (Y_1,\ldots,Y_n)$ has to be communicated via a noiseless digital channel whose capacity is $R$ bits per sample. This situation, shown in Figure~\ref{fig:rate_limited_learning}, may arise, for example, in remote sensing, where the $X_i$'s are the locations of the sensors and the $Y_i$'s are the measurements of the sensors having the form $f_0(X_i) + Z_i$, where $\map{f_0}{\cX}{[0,1]}$ is some unknown function and the $Z_i$'s are i.i.d. zero-mean Gaussian random variables with variance $\sigma^2$. Assuming that the sensors are dispersed at random over some bounded spatial region $\cX$ and the location of each sensor is known following its deployment, the task of the sensor array is to deliver, over a rate-limited channel, an approximation $\wh{Y}^n$ of the measurement vector $Y^n = (Y_1,\ldots,Y_n)$ to some central location, where the vector $X^n$ of the sensor locations and the compressed version $\wh{Y}^n$ of the sensor measurements will be fed into a learner that will approximate $f_0$ by some function $\wh{f}_n(X^n,\wh{Y}^n,\cdot)$ from a given hypothesis class $\cF$.

In this paper, we establish information-theoretic upper bounds on the achievable generalization error in this setting. In particular, we relate the problem of agnostic learning under (partial) rate constraints to conditional rate-distortion theory \cite[Section~6.1]{Ber71}, \cite{Gra72}, \cite[Appendix~A]{Wyn78}, which is concerned with lossy source coding in the presence of side information both at the encoder and at the decoder. In the set-up shown in Figure~\ref{fig:rate_limited_learning}, the input part $X^n = (X_1,\ldots,X_n)$ of the training sequence, which is available both to the encoder and to the decoder (hence to the learner), plays the role of the side information, while the output part $Y^n = (Y_1,\ldots,Y_n)$ is to be coded using a lossy source code operating at the rate of $R$ bits per symbol. Furthermore, because the distribution of $(X,Y)$ is known only to be a member of some family $\cP$, the lossy codes must be robust in the presence of this uncertainty.

Let us formally state the problem. Let $\cP,\cF,\ell$ be given. A scheme for agnostic learning under partial rate constraints (from now on, simply a {\em scheme}) operating at rate $R$ is specified by a sequence of triples $\{ (e_n, d_n, \wh{f}_n) \}^\infty_{n=1}$, where $\map{e_n}{\cX^n \times \cY^n}{\{1,\ldots,2^{nR}\}}$ is the encoder, $\map{d_n}{\cX^n \times \{1,\ldots,2^{nR}\}}{\cY^n}$ is the decoder, and
$\map{\wh{f}_n}{\cX^n \times \cY^n}{\cF}$ is the learner. We shall often abuse notation and let $\wh{f}_n$ denote also the function $\wh{f}_n(X^n,\wh{Y}^n,\cdot)$. For each $n$, the output of the learner is a hypothesis $\wh{f}_n(X^n,\wh{Y}^n,\cdot) \in \cF$, where $\wh{Y}^n = d_n(X^n,e_n(X^n,Y^n))$ is the reproduction of $Y^n$ given the side information $X^n$. For any $P \in \cP$, the main object of interest associated with the scheme is the generalization error
$$
L(\wh{f}_n,P) \deq \E \Big[ \ell \big( \wh{f}_n(X^n,\wh{Y}^n,X),Y \big) \Big| X^n, Y^n \Big],
$$
where $(X,Y) \sim P$ is assumed independent of $\{(X_i,Y_i)\}^n_{i=1}$ (to keep the notation simple, we suppress the dependence of the generalization error on the encoder and the decoder). In particular, we are interested in the achievable values of the asymptotic expected excess risk. We say that a pair $(R,\Delta)$ is {\em achievable for $(\cF,\cP,\ell)$} if there exists a scheme $\{(e_n,d_n,\wh{f}_n)\}^\infty_{n=1}$ operating at rate $R$, such that
$$
\limsup_{n \to \infty} \E L(\wh{f}_n,P) \le L^*(\cF,P) + \Delta
%\label{eq:achievability}
$$
for every $P \in \cP$. After listing the basic assumptions in Sec.~\ref{sec:assumptions}, we derive in Sec.~\ref{sec:results} sufficient conditions for $(R,\Delta)$ to be achievable. We then apply our results to the setting of nonparametric regression in Sec.~\ref{sec:regression}. Discussion of results and an outline of future directions are given in Sec.~\ref{sec:discussion}.

\subsection{Related work}

Previously, the problem of statistical estimation from compressed data was considered by Zhang and Berger \cite{ZhaBer88}, Ahlswede and Burnashev \cite{AhlBur90} and Han and Amari \cite{HanAma98} from the viewpoint of multiterminal information theory. In these papers, the underlying family of distributions of $(X,Y)$ is parametric, i.e., of the form $\cP = \{P_\theta\}_{\theta \in \Theta}$, where $\Theta$ is a subset of $\R^k$ for some finite $k$, and one wishes to estimate the ``true" parameter $\theta^*$. The i.i.d.~observations $\{(X_i,Y_i)\}^n_{i=1}$ are drawn from $P_{\theta^*}$, and the input part $X^n$ is communicated to the statistician at some rate $R_1$, while the output part $Y^n$ is communicated at some rate $R_2$. The present work generalizes to the nonparametric setting the case considered by Ahlswede and Burnashev \cite{AhlBur90}, namely when $R_1 = \infty$. To the best of the author's knowledge, this paper is the first to consider the problem of nonparametric learning from compressed observations with side information.

\section{Assumptions}
\label{sec:assumptions}

We begin by stating some basic assumptions on $\cF$, $\cP$ and $\ell$. Additional assumptions will be listed in the sequel as needed.

The input space $\cX$ is taken to be a measurable subset of $\R^d$, while the output space is either a finite set (as in classification) or the set of reals $\R$ (as in regression or function estimation). We assume throughout that the family $\cP$ of distributions on $\cX \times \cY$ is such that the mutual information $I(X;Y) < \infty$ for every $P \in \cP$. All information-theoretic quantities will be in bits, unless specified otherwise.

\begin{comment}
All finite sets are assumed equipped with the $\sigma$-algebras of all their subsets, while the Euclidean spaces are assumed equipped with their Borel $\sigma$-algebras. All random variables taking values in continuous spaces are assumed to have finite differential entropies. (Recall that the differential entropy of a random variable $U$ taking values in $\R^d$ and possessing a probability density function $p(u)$ with respect to the Lebesgue measure is defined as
\begin{equation}
h(U) \deq - \int_{\R^d}p(u)\log p(u) du;
\label{eq:differential_entropy}
\end{equation} 
see, e.g., Chapter~9 of \cite{CovTho91}.) All functions are assumed to be measurable with respect to appropriate $\sigma$-algebras.
\end{comment}

We assume that there exists a learning algorithm which generalizes optimally in the absence of any rate constraints. Therefore, our standing assumption on $(\cF,\cP,\ell)$ will be that the induced function class $\cL_\cF = \{\ell_f : f \in \cF\}$, where $\ell_f(z) \deq \ell(f(x),y)$ for all $z = (x,y) \in \cZ$, satisfies the {\em uniform law of large numbers} (ULLN) for every $P \in \cP$, i.e.,
\begin{equation}
\sup_{f \in \cF} \left| \frac{1}{n}\sum^n_{i=1} \ell_f(Z_i) - \E \ell_f(Z) \right| \to 0, \qquad \rm{a.s.}
\label{eq:ULLN}
\end{equation}
where $Z,Z_1,Z_2,\ldots$ are i.i.d. according to $P$. Eq.~(\ref{eq:ULLN}) implies that, for any sequence $\{f_n\} \subset \cF$,
$$
\left| \frac{1}{n}\sum^n_{i=1} \ell_{f_n}(Z_i) - \E\ell_{f_n}(Z)\right| \to 0, \qquad \rm{a.s.}
$$
This holds even in the case when each $f_n$ is random, i.e., $f_n(\cdot) = f_n(Z^n,\cdot)$. The ULLN is a standard ingredient in  proofs of consistency of learning algorithms: if $(\cF,\cP,\ell)$ are such that (\ref{eq:ULLN}) holds, then the {\em Empirical Risk Minimization} algorithm (ERM), given by
$$
\wh{f}_n = \argmin_{f \in \cF} \frac{1}{n}\sum^n_{i=1} \ell_f(Z_i),
%\label{eq:ERM}
$$
is PAC in the sense of (\ref{eq:universal_consistency}) \cite[Theorem~3.2]{Vid03}.

Next, we assume that the loss function $\ell$ has the following ``generalized Lipschitz" property: there exists a concave, continuous function $\map{\eta}{\R^+}{\R^+}$, such that for all $f \in \cF$, $x \in \cX$ and $u,u' \in \cY$
\begin{equation}
\left| \ell(f(x),u) - \ell(f(x),u') \right| \le  \eta(\ell(u,u')).
\label{eq:loss_function_smoothness}
\end{equation}
This holds, for example, in the following cases:
\begin{itemize}
\item Suppose that $\ell$ is a metric on $\cY$. Then, by the triangle inequality we have $\ell(y,u) \le \ell(y,u') + \ell(u',u)$ for all $y,u,u' \in \cY$, so (\ref{eq:loss_function_smoothness}) holds with $\eta(t) = t$.
\item Suppose that $\cY = [0,1]$ and $\ell(u,u') = |u-u'|^p$ for some $p \ge 1$. Then one can show that
$$
\left| \ell(f(x),u) - \ell(f(x),u') \right| \le p | u - u'|
$$
for all $\map{f}{\cX}{\cY}$, $x \in \cX$ and $u,u' \in \cY$, so (\ref{eq:loss_function_smoothness}) holds with $\eta(t) = p t^{1/p}$.
\end{itemize}

Finally, we need to pose some assumptions on the metric structure of the class $\cP$ with respect to the {\em variational distance} \cite[Sec.~5.2]{Gra90a}, which for any two probability distributions $P_1,P_2$ on a measurable space $(\cZ,\cA)$ is defined by
$$
d_V(P_1,P_2) \deq 2\sup_{A \in \cA} |P_1(A) - P_2(A)|.
%\label{eq:variational_distance}
$$
A finite set $\{P_1,\ldots,P_M\} \subset \cP$ is called an {\em (internal) $\epsilon$-net} for $\cP$ with respect to $d_V$ if
$$
\sup_{P \in \cP} \min_{1 \le m \le M} d_V(P,P_m) \le \epsilon.
%\label{eq:epsilon_net}
$$
The cardinality of a minimal $\epsilon$-net, denoted by $N(\epsilon,\cP)$, is called the {\em $\epsilon$-covering number} of $\cP$ w.r.t. $d_V$, and the {\em Kolmogorov $\epsilon$-entropy} of $\cP$ is defined as $H(\epsilon,\cP) \deq \log N(\epsilon,\cP)$ \cite{KolTih61}. We assume that the class $\cP$ satisfies {\em Dobrushin's entropy condition} \cite{Dob70}, i.e., for every $c > 0$
\begin{equation}
\lim_{\epsilon \to 0} \frac{H(\epsilon,\cP)}{2^{c/\epsilon}} = 0.
\label{eq:dobrushin_condition}
\end{equation}
This condition is satisfied, for example, in the following cases: (1) $\cX$ and $\cY$ are both finite sets; (2) $\cP$ is a finite family; (3) $\cZ$ is a compact subset of a Euclidean space, and all $P \in \cP$ are absolutely continuous with densities satisfying a uniform Lipschitz condition \cite{KolTih61,Dob70}. 

\section{The results}
\label{sec:results}

To state our results we shall need some notions from conditional rate-distortion theory \cite[Sec.~6.1]{Ber71}, \cite{Gra72}, \cite[Appendix~A]{Wyn78}. Fix some $P \in \cP$. Given a pair $(X,Y) \sim P$ and a nonnegative real number $D$, define the set $\cM(D)$ to consist of all $\cY$-valued random variables $\wh{Y}$ jointly distributed with $(X,Y)$ and satisfying the constraint $\E \ell(Y,\hat{Y}) \le D$, where the expectation is taken with respect to the joint distribution of $X,Y,\wh{Y}$. Then the {\em conditional rate-distortion function} of $Y$ given $X$ w.r.t.~$P$ is defined by
$$
R_{Y|X}(D,P) \deq \inf \left\{ I(Y; \wh{Y} | X) : \wh{Y} \in \cM(D) \right\},
%\label{eq:conditional_RDF}
$$
where $I(Y; \wh{Y} | X)$ is the conditional mutual information between $Y$ and $\wh{Y}$ given $X$. Our assumption that $I(X;Y) < \infty$ ensures the existence of $R_{Y|X}(D,P)$ \cite{Wyn78}. In operational terms, $R_{Y|X}(D,P)$ is the minimum number of bits needed to describe $Y$ with expected distortion of at most $D$ given perfect observation of a correlated random variable $X$ (the side information) when $(X,Y) \sim P$. As a function of $D$, $R_{Y|X}(D,P)$ is convex and strictly decreasing everywhere it is finite, hence it is invertible. The inverse function is called the {\em conditional distortion-rate function} of $Y$ given $X$ and is denoted by $D_{Y|X}(R,P)$. Finally, let
$$
\bbD_{Y|X}(R,\cP) \deq \sup_{P \in \cP} D_{Y|X}(R,P).
%\label{eq:sup_conditional_DRF}
$$
We assume that $\bbD_{Y|X}(R,\cP) < \infty$ for all $R \ge 0$.

We shall also need the following lemma, which can be proved by a straightforward extension of Dobrushin's random coding argument from \cite{Dob70} to the case of side information available to the encoder and to the decoder:

\begin{lemma}\label{lm:robust_codes} Let $\cP$ satisfy Dobrushin's entropy condition (\ref{eq:dobrushin_condition}). Assume that the loss function $\ell$ either is bounded or satisfies a uniform moment condition
\begin{equation}
\sup_{P \in \cP} \E[\ell(Y,y_0)^{1+\delta}] < \infty
\label{eq:moment_condition}
\end{equation}
for some $\delta > 0$ with respect to some fixed reference letter $y_0 \in \cY$. Then for every rate $R \ge 0$ there exists a sequence $\{(e_n,d_n)\}^\infty_{n=1}$ of encoders $\map{e_n}{\cX^n \times \cY^n}{\{1,\ldots,2^{nR}\}}$ and decoders $\map{d_n}{\cX^n \times \{1,\ldots,2^{nR}\}}{\cY^n}$, such that
$$
\limsup_{n \to \infty} \sup_{P \in \cP} \E \ell_n(Y^n,\wh{Y}^n) \le \bbD_{Y|X}(R,\cP),
$$
where $\wh{Y}^n = d_n(X^n,e_n(X^n,Y^n))$ and $\ell_n(Y^n,\wh{Y}^n) = n^{-1}\sum^n_{i=1} \ell(Y_i,\wh{Y}_i)$ is the normalized cumulative loss between $Y^n$ and $\wh{Y}^n$.
\end{lemma}\vspace{5pt}

\noindent Our main result can then be stated as follows:

\begin{theorem}\label{thm:upper_bound}
Under the stated assumptions, for any $R \ge 0$ there exists a scheme $\{ (e_n,d_n,\wh{f}_n)\}$ operating at rate $R$, such that
$$
\limsup_{n \to \infty} \E L(\wh{f}_n,P) \le L^*(\cF,P) + 2\eta(\bbD_{Y|X}(R,\cP)).
$$
Thus, $(R,2\eta(\bbD_{Y|X}(R,\cP)))$ is achievable for every $R \ge 0$.
\end{theorem}\vspace{5pt}

\begin{comment}
\begin{eqnarray}
nR &\ge& H(J|X^n) \\
&\ge& H(\wh{Y}^n|X^n) \\
&\ge& I(\wh{Y}^n; Y^n | X^n) \\
&=& H(Y^n | X^n) - H(Y^n | X^n, \wh{Y}^n) \\
&=& H(Y^n | X^n) - H(Y^n | X^n, \wh{Y}^n, W^n) \\
&=& \sum^n_{i=1} [H(Y_i | X_i) - H(Y_i | X^n, \wh{Y}^n, W^n, Y^{i-1} ) ] \\
&\ge& \sum^n_{i=1} [H(Y_i | X_i) - H(Y_i | X_i, W_i)] \\
&=& \sum^n_{i=1} I(Y_i; W_i | X_i) \\
&\ge& \sum^n_{i=1} R_{Y|X} \left(\E \ell(W_i,Y_i) \right) \\
&=& n\sum^n_{i=1} \frac{1}{n} R_{Y|X} \left(\E \ell(W_i,Y_i) \right) \\
&\ge& nR_{Y|X}\left(\frac{1}{n}\sum^n_{i=1} \E \ell(W_i,Y_i) \right) \\
&=& nR_{Y|X}\left(\E \wh{L}(\widehat{f}_n)\right).
\end{eqnarray}
Thus, $\E \wh{L}(\widehat{f}_n) \ge D_{Y|X}(R)$.
\end{comment}

\begin{proof} Given $n$, $Z^n \in \cZ^n$ and $f \in \cF$, define the {\em empirical risk}
$$
\wh{L}_{Z^n}(f) \deq \frac{1}{n} \sum^n_{i=1} \ell_f(Z^n)$$
and the minimum empirical risk
$$
\wh{L}^*_{Z^n}(\cF) \deq \Inf_{f \in \cF} \wh{L}_{Z^n}(f).$$
We shall write $\wh{L}_{X^n,Y^n}(f)$ and $\wh{L}^*_{X^n,Y^n}(\cF)$ whenever we need to emphasize separately the roles of $X^n$ and $Y^n$.

Suppose that the encoder $e_n$ and the decoder $d_n$ are given. Let $\wh{Y}^n$ denote the reproduction of $Y^n$ given the side information $X^n$, i.e., $\wh{Y}^n = d_n(X^n,e_n(X^n,Y^n))$. We then define our learner $\wh{f}_n$ by
\begin{equation}
\wh{f}_n = \argmin_{f \in \cF} \wh{L}_{X^n,\wh{Y}^n}(f).
\label{eq:quantized_ERM}
\end{equation}
In other words, having received the side information $X^n$ and the reproduction $\wh{Y}^n$, the learner performs ERM over $\cF$ on $\{(X_i,\wh{Y}_i)\}^n_{i=1}$. Using the property (\ref{eq:loss_function_smoothness}) of the loss function $\ell$ and the concavity of $\eta$, we have the following estimate:
\begin{eqnarray}
&& \sup_{f \in \cF} \big| \wh{L}_{X^n,Y^n}(f)  - \wh{L}_{X^n,\wh{Y}^n}(f) \big| \nonumber \\
&& \quad \le \sup_{f \in \cF} \frac{1}{n}\sum^n_{i=1} \big| \ell(f(X_i),Y_i) - \ell(f(X_i),\wh{Y}_i) \big| \nonumber \\
&& \quad \le \frac{1}{n}\sum^n_{i=1} \eta (\ell(Y_i,\wh{Y}_i)) \nonumber \\
&& \quad \le  \eta \big( \ell_n(Y^n,\wh{Y}^n )\big). \label{eq:loss_approximation}
\end{eqnarray}
In particular, this implies that
\begin{equation}
\big|\wh{L}_{X^n,Y^n}(\wh{f}_n) - \wh{L}_{X^n,\wh{Y}^n}(\wh{f}_n) \big| \le \eta\big( \ell_n(Y^n, \wh{Y}^n) \big)
\label{eq:approximation_bound_1}
\end{equation}
and
\begin{equation}
\big| \wh{L}^*_{X^n,Y^n}(\cF) - \wh{L}^*_{X^n,\wh{Y}^n}(\cF) \big| \le \eta\big( \ell_n(Y^n,\wh{Y}^n) \big).
\label{eq:approximation_bound_2}
\end{equation}
We then have
\begin{eqnarray*}
\wh{L}_{X^n,Y^n}(\wh{f}_n) &\stackrel{{\rm (a)}}{\le}& \wh{L}_{X^n,\wh{Y}^n}(\wh{f}_n) + \eta \big( \ell_n(Y^n, \wh{Y}^n) \big) \\
&\stackrel{{\rm (b)}}{=}& \wh{L}^*_{X^n,\wh{Y}^n}(\cF) + \eta \big( \ell_n(Y^n,\wh{Y}^n) \big) \\
&\stackrel{{\rm (c)}}{\le}& \wh{L}^*_{X^n,Y^n}(\cF) + 2\eta\big( \ell_n(Y^n,\wh{Y}^n) \big), 
\end{eqnarray*}
where (a) follows from (\ref{eq:approximation_bound_1}), (b) from the definition of $\wh{f}_n$, and (c) from (\ref{eq:approximation_bound_2}). Suppose that the data are distributed according to a particular $P \in \cP$. Taking expectations and using the concavity of $\eta$ and Jensen's inequality, we obtain
$$
\E \wh{L}_{Z^n}(\wh{f}_n) \le \E \wh{L}^*_{Z^n}(\cF) + 2\eta \big( \E \ell_n(Y^n,\wh{Y}^n) \big).
$$
Using this bound and the continuity of $\eta$, we can write
\begin{eqnarray}
&& \limsup_{n \to \infty} \E L(\wh{f}_n,P) - L^*(\cF,P) \nonumber \\
&& \qquad \le \lim_{n \to \infty} \E \big [ L(\wh{f}_n,P) - \wh{L}_{Z^n}(\wh{f}_n) \big ] \nonumber \\
&& \qquad\quad  + \lim_{n \to \infty} \E \big [ \wh{L}^*_{Z^n}(\cF) - L^*(\cF,P)\big] \nonumber \\
&& \qquad\quad +  2\eta \Big( \limsup_{n \to \infty} \E \ell_n(Y^n,\wh{Y}^n) \Big). \label{eq:excess_loss_bound}
\end{eqnarray}
The two leading terms on the right-hand side of this inequality are zero by the ULLN. Moreover, given $R$, Lemma~\ref{lm:robust_codes} asserts the existence of a sequence $\{(e_n,d_n)\}^\infty_{n=1}$ of encoders $\map{e_n}{\cX^n \times \cY^n}{\{1,\ldots,2^{nR}\}}$ and decoders $\map{d_n}{\cX^n \times \{1,\ldots,2^{nR}\}}{\cY^n}$, such that
$$
\limsup_{n \to \infty} \E\ell_n(Y^n,\wh{Y}^n) \le \bbD_{Y|X}(R,\cP), \qquad \forall P \in \cP.
$$
Substitution of this into (\ref{eq:excess_loss_bound}) proves the theorem.
\end{proof}

\begin{corollary} All pairs $(R,\Delta)$ with $\Delta \ge 2\eta(\bbD_{Y|X}(R,\cP))$ are achievable.
\end{corollary}

\begin{remark}\label{rem:lower_bound} In the Appendix, we show that a corresponding lower bound derived by the usual methods for proving converses in lossy source coding is strictly weaker than the ``obvious" lower bound based on the observation that $\E L(\wh{f}_n,P) \ge L^*(\cF,P)$ for any $\wh{f}_n$. It may be possible to obtain nontrivial lower bounds in the minimax setting, which we leave for future work (see also Sec.~\ref{sec:discussion}).
\end{remark}

\begin{remark}\label{rem:finite_sample_bound} Under some technical conditions on the function class $\{ \ell_f: f \in \cF\}$ (see, e.g., \cite{Men03}), one can show that
$$
\E \sup_{f \in \cF} \Big| \wh{L}_{Z^n}(f) - L(f,P) \Big| \le C/\sqrt{n}, \qquad \forall P \in \cP
$$
for some constant $C$ that depends on $\cF,\ell$. Using this fact and the same bounding method that led to Eq.~(\ref{eq:excess_loss_bound}), but without taking the limit superior, we can get the following finite-sample bound for every scheme $\{(e_n,d_n,\wh{f}_n)\}^\infty_{n=1}$ with $\wh{f}_n$ given by (\ref{eq:quantized_ERM}) and {\em arbitrary} $e_n,d_n$:
$$
 \E L(\wh{f}_n,P) \le L^*(\cF,P) + 2\eta\big(\E \ell_n(Y^n,\wh{Y}^n) \big) + C'/\sqrt{n},
$$
where $C' = 2C$.
\end{remark}\vspace{5pt}

\noindent The following theorem shows that we can replace condition (\ref{eq:loss_function_smoothness}) with the requirement that $\ell$ be a power of a metric:

\begin{theorem}\label{thm:metric_power} Suppose that the loss function $\ell$ is of the form $\ell(y,u) = d(y,u)^r$ for some $r \ge 1$, where $d$ is a metric on $\cY$. Then for any rate $R \ge 0$ the scheme constructed in the proof of Theorem~\ref{thm:upper_bound} is such that
$$
\limsup_{n \to \infty} \E \Big[L(\wh{f}_n,P)^{1/r}\Big] \le L^*(\cF,P)^{1/r} + 2 \bbD_{Y|X}(R,\cP)^{1/r}
$$
holds for every $P \in \cP$.\end{theorem}\vspace{5pt}

\begin{proof} We proceed essentially along the same lines as in the proof of Theorem~\ref{thm:upper_bound}, except that the bound (\ref{eq:loss_approximation}) is replaced with an argument based on Minkowski's inequality to yield
$$
\E\Big[\wh{L}_{Z^n}(\wh{f}_n)^{1/r}\Big] \le \E \Big[\wh{L}^*_{Z^n}(\cF)^{1/r}\Big] + 2\Big(\E \ell_n(Y^n,\wh{Y}^n) \Big)^{1/r}.$$
The rest is immediate using the ULLN as well as concavity and continuity of $t \mapsto t^{1/r}$ for $t \ge 0$. \end{proof}

\section{Example: nonparametric regression}
\label{sec:regression}

As an example, let us consider the setting of nonparametric regression. Let $\cX$ be a compact subset of $\R^d$ and $\cY = \R$. The training data are of the form
\begin{equation}
Y_i = f_0(X_i) + Z_i, \qquad 1 \le i \le n
\label{eq:nonparametric_regression}
\end{equation}
where the regression function $f_0$ belongs to some specified class $\cF$ of functions from $\cX$ into $[0,1]$, the $X_i$'s are i.i.d.~random variables drawn from the uniform distribution on $\cX$, and the $Z_i$'s are i.i.d.~zero-mean normal random variables with variance $\sigma^2$, independent of $X^n$. We take $\ell(y,u) = |y-u|^2$, the squared loss. Note that $\ell$ satisfies the condition of Theorem~\ref{thm:metric_power} with $r = 2$.

Because $f_0$ is unknown, we take as the underlying family $\cP$ the class of all absolutely continuous distributions with densities of the form $p_f(x,y) = V^{-1} \cN(y; f(x),\sigma^2)$, $f \in \cF$, where $V$ is the volume of $\cX$ and $\cN(y; f(x),\sigma^2)$ is the one-dimensional normal density with mean $f(x)$ and variance $\sigma^2$. Because the functions in $\cF$ are bounded between $0$ and $1$, it is easy to show that the uniform moment condition (\ref{eq:moment_condition}) of Lemma~\ref{lm:robust_codes} is satisfied with $\delta = 1$ and $y_0 = 0$.

We suppose that $\ell$ and $\cF$ are such that the function class $\cL_\cF$ satisfies the ULLN.\footnote{See Gy\"orfi et al.~\cite{GKKW02} for a detailed exposition of the various conditions when this is true.} Let $Q$ denote the uniform distribution on $\cX$ and for any square-integrable function $f$ on $\cX$ define the $L_2$ norm by
$$
\| f \|^2_{2,Q} \deq \int_\cX f^2(x) dQ(x) \equiv \frac{1}{V}\int_\cX f^2(x) dx.
$$
Let us denote by $N_{2,Q}(\epsilon,\cF)$ the $\epsilon$-covering number of $\cF$ w.r.t. $\| \cdot \|_{2,Q}$, i.e., the smallest number $M$ such that there exist $M$ functions $\{ f_m \}^M_{m=1}$ in $\cF$ satisfying
$$
\sup_{f \in \cF} \min_{1 \le m \le M} \| f - f_m \|_{2,Q} \le \epsilon.
$$
We assume that $\cF$ is such that for every $c > 0$
\begin{equation}
\lim_{\epsilon \to 0} \frac{\log N_{2,Q}(\epsilon,\cF)}{2^{c/\epsilon}} = 0.
\label{eq:function_covering_condition}
\end{equation}
This condition holds, for example, if the functions in $\cF$ are uniformly Lipschitz or if $\cX$ is a bounded interval in $\R$ and $\cF$ consists of functions satisfying a Sobolev-type condition \cite{KolTih61}.

\begin{lemma} If $\cF$ satisfies (\ref{eq:function_covering_condition}), then $\cP$ satisfies Dobrushin's entropy condition (\ref{eq:dobrushin_condition}).\end{lemma}\vspace{5pt}

\begin{proof} Given $f \in \cF$, let $P_f$ denote the distribution with the density $p_f$. It is straightforward to show that
$$
I(P_f \| P_g) = \frac{1}{2\sigma^2} \| f - g \|^2_{2,Q}, \qquad \forall f,g \in \cF
%\label{eq:relative_entropy}
$$
where $I(\cdot \| \cdot)$ is the relative entropy (information divergence) between two probability distributions, in nats. Using Pinsker's inequality $d_V(P_1,P_2) \le \sqrt{2 I(P_1 \| P_2)}$ \cite[Lemma~5.2.8]{Gra90a}, we get
\begin{equation}
d_V(P_f \| P_g) \le \frac{1}{\sigma} \| f - g \|_{2,Q}, \qquad \forall f,g \in \cF.
\label{eq:dvar_bound}
\end{equation}
Given $\epsilon > 0$, let $\{f_m \}^M_{m=1} \subset \cF$ be a $\sigma\epsilon$-net for $\cF$ w.r.t.~$\|\cdot\|_{2,Q}$. Then from (\ref{eq:dvar_bound}) it follows that
$$
\sup_{f \in \cF} \min_{1 \le m \le M}d_V(P_f,P_{f_m}) \le  \sup_{f \in \cF} \min_{1 \le m \le M} \frac{\| f - f_m \|_{2,Q}}{\sigma} \le \epsilon,
$$
i.e., $\{P_{f_m}\}^M_{m=1}$ is an $\epsilon$-net for $\cP$ w.r.t.~$d_V$. This implies, in particular, that $N(\epsilon,\cP) \le N_{2,Q}(\sigma\epsilon,\cF)$ for every $\epsilon > 0$. This, together with (\ref{eq:function_covering_condition}), proves the lemma.\end{proof}

\begin{lemma}\label{lm:sup_DRF}
For any $R \ge 0$, $\bbD_{Y|X}(R,\cP) = \sigma^2 2^{-2R}$.
\end{lemma}\vspace{5pt}

\begin{proof} Fix some $f \in \cF$ and consider a pair $(X,Y) \sim P_f$. Then $Y = f(X) + Z$, where $Z \sim {\rm Normal}(0,\sigma^2)$ is independent of $X$. Because $\ell$ is a difference distortion measure, Theorem~7 of \cite{Gra72} says that, for any measurable function $\map{\psi}{\cX}{\cY}$,
$$
D_{Y|X}(R,P_f) = D_{Y - \psi(X)|X}(R,P_{f-\psi}),
$$
where $P_{f-\psi}$ is the distribution of
$$Y - \psi(X) \equiv f(X) - \psi(X) + Z;$$
furthermore, if $Y - \psi(X)$ is independent of $X$, then $D_{Y|X}(R,P_f) = D_{Y - \psi(X)}(R)$, the (unconditional) distortion-rate function of $Y - \psi(X)$. Taking $\psi = f$, we get $D_{Y|X}(R,P_f) = D(R,\sigma^2)$, the distortion-rate function of a memoryless Gaussian source with variance $\sigma^2$ w.r.t.~squared error loss, which is equal to $\sigma^2 2^{-2R}$ \cite[Theorem~9.3.2]{Ber71}. Hence $D_{Y|X}(R,P_f)$ is independent of $f$. Taking the supremum over $\cF$ finishes the proof.\end{proof}\vspace{5pt}

\noindent Now we can state and prove the main result of this section:

\begin{theorem} Consider the regression setting of (\ref{eq:nonparametric_regression}). Under the stated assumptions, for any $R \ge 0$ there exists a scheme $\{ (e_n,d_n,\wh{f}_n )\}^\infty_{n=1}$, such that
\begin{equation}
\limsup_{n \to \infty}  \E \left[ L(\wh{f}_n,P_f)^{1/2} \right] \le \sigma(1 + 2^{-R+1}) \label{eq:regression_bound_2}
\end{equation}
holds for every $f \in \cF$.
\end{theorem}\vspace{5pt}

\begin{proof} As follows from the above, the triple $(\cF,\cP,\ell)$ satisfies all the assumptions of Theorem~\ref{thm:metric_power}. Therefore for any $R \ge 0$ there exists a scheme $\{ (e_n,d_n,\wh{f}_n ) \}^\infty_{n=1}$ operating at rate $R$, such that
\begin{equation}
\limsup_{n \to \infty} \E  \left[ L(\wh{f}_n,P_f)^{1/2} \right] \le L^*(\cF,P_f)^{1/2} + 2^{-R+1}\sigma, \label{eq:regression_bound_2a}
\end{equation}
holds for every $f \in \cF$ (we have also used Lemma~\ref{lm:sup_DRF}). It is not hard to show that
$$
L(g,P_f) = \| f - g \|^2_{2,Q} + \sigma^2, \qquad \forall f,g \in \cF,
$$
whence it follows that $L^*(\cF,P_f) = \sigma^2$ for every $f \in \cF$. Substituting this into (\ref{eq:regression_bound_2a}), we get (\ref{eq:regression_bound_2}). 
\end{proof}

\section{Discussion and future work}
\label{sec:discussion}

We have derived information-theoretic bounds on the achievable generalization error in learning from compressed data (with side information). There is a close relationship between this problem and the theory of robust lossy source coding with side information at the encoder and the decoder. A major difference between this setting and the usual setting of learning theory is that the techniques are no longer {\em distribution-free} because restrictions must be placed on the underlying family of distributions in order to guarantee the existence of a suitable source code. The theory was applied to the problem of nonparametric regression in Gaussian noise, where we have shown that the penalty incurred for using compressed observations decays exponentially with the rate.

We have proved Theorems~\ref{thm:upper_bound} and \ref{thm:metric_power} by adopting ERM as our learning algorithm and optimizing the source code to deliver the best possible reconstruction of the training data. In effect, this imposes a {\em separation structure} between learning and source coding. While this ``modular" approach is simplistic (clearly, additional performance gains could be attained by designing the encoder, the decoder and the learner jointly), it may be justified in such applications as remote sensing. For instance, if the source code and the learner were designed jointly, then any change made to the hypothesis class (say, if we decided to replace the currently used hypothesis class with another based on tracking the prior performance of the network) might call for a complete redesign of the source code and the sensor network, which may be a costly step. With the modular approach, no such redesign is necessary: one merely makes the necessary adjustments in the learning algorithm, while the sensor network continues to operate as before.

Let us close by sketching some directions for future work. First of all, it would be of interest to derive information-theoretic lower bounds on the generalization performance of rate-constrained learning algorithms. In particular, just as Ahlswede and Burnashev had done in the parametric case \cite{AhlBur90}, we could study the asymptotics of the {\em ninimax excess risk}
$$
\delta_n(R) \deq \inf_{(e_n,d_n,\wh{f}_n)} \sup_{P \in \cP} \Big[\E L(\wh{f}_n,P) - L^*(\cF,P) \Big],
$$
where the infimum is over all encoders, decoders and learners operating on a length-$n$ training sequence at rate $R$. Secondly, we could dispense with the assumption that the learner has perfect observation of the input part of the training sample, in analogy to the situation dealt with by Zhang and Berger \cite{ZhaBer88}. Finally, keeping in mind the motivating example of sensor networks, it would be useful to replace the block coding approach used here with an efficient distributed scheme.

\section*{Acknowledgments}
\noindent Discussions with Todd Coleman are gratefully acknowledged. This work was supported by the Beckman Fellowship.

\appendix
\renewcommand{\theequation}{A.\arabic{equation}}
\setcounter{equation}{0}

Let us assume for simplicity that $\cP$ is a singleton, $\cP = \{P\}$, and that $\cY$ is a finite set. Consider a scheme $\{(e_n,d_n,\wh{f}_n)\}$ operating at rate $R$. Fix $n$ and define the $n$-tuple $W^n$ via
$$
W_i \deq \wh{f}_n(X^n,\wh{Y}^n,X_i), \qquad 1 \le i \le n.
$$
Also, let $J = e_n(X^n,Y^n)$. Then we can write
\begin{eqnarray}
nR &\ge& H(J|X^n) \nonumber \\
&\ge& H(\wh{Y}^n|X^n) \nonumber \\
&\ge& I(\wh{Y}^n; Y^n | X^n) \nonumber \\
&=& H(Y^n | X^n) - H(Y^n | X^n, \wh{Y}^n) \nonumber \\
&=& H(Y^n | X^n) - H(Y^n | X^n, \wh{Y}^n, W^n) \label{eq:fctn}
\end{eqnarray}
\begin{eqnarray*}
&=& \sum^n_{i=1} [H(Y_i | X_i) - H(Y_i | X^n, \wh{Y}^n, W^n, Y^{i-1} ) ] \nonumber \\
&\ge& \sum^n_{i=1} [H(Y_i | X_i) - H(Y_i | X_i, W_i)] \nonumber \\
&=& \sum^n_{i=1} I(Y_i; W_i | X_i) \nonumber \\
&\ge& \sum^n_{i=1} R_{Y|X} (\E \ell(W_i,Y_i),P) \nonumber \\
&\ge& nR_{Y|X}(\E \ell_n(W^n,Y^n),P),\nonumber
\end{eqnarray*}
where (\ref{eq:fctn}) follows from the fact that $W^n$ is a function of $\wh{Y}^n$ and $X^n$. The remaining steps follow from standard information-theoretic identities and from convexity. Therefore,
$$
\liminf_{n \to \infty} \E\ell_n(W^n,Y^n) \ge D_{Y|X}(R,P).
$$
Because $\E L(\wh{f}_n,P) = \E \ell_n(W^n,Y^n) + o(1)$ by the ULLN,
\begin{equation}
\liminf_{n \to \infty} \E L(\wh{f}_n,P) \ge D_{Y|X}(R,P).
\label{eq:info_lower_bound}
\end{equation}
Now, given any $f \in \cF$, we can interpret $f(X)$ as a zero-rate approximation of $Y$ (using only the side information $X$), so $L(f,P) \ge D_{Y|X}(0,P) \ge D_{Y|X}(R,P)$ for any $R \ge 0$. In particular, $L^*(\cF,P) \ge D_{Y|X}(R,P)$ for all $R$, and
$$
\liminf_{n \to \infty} \E L(\wh{f}_n,P) \ge L^*(\cF,P) \ge D_{Y|X}(R,P)
$$
for all $R$. Thus, the information-theoretic lower bound (\ref{eq:info_lower_bound}) is weaker than the bound $\Liminf_{n \to \infty} \E L(\wh{f}_n,P) \ge L^*(\cF,P)$.

\bibliography{rate_constrained_learning}

\end{document}